%Paper: hep-ph/9501269
%From: FTLIPKIN@WEIZMANN.weizmann.ac.il
%Date: 11 Jan 95 15:49 +02

%macropackage=phyzzx
\vsize=7.5in
\hsize=6.6in
\hfuzz=20pt%this is because one equation is a bit wide.
\tolerance 10000

\baselineskip 12pt plus 1pt minus 1pt

\def\){]}
\def\({[}

\def\rjustline#1{\line{\hss#1}}

\rjustline{WIS-95/1/Jan-PH}
\rjustline{TAUP 2225-95}

\centerline{\bf
Theories of Non-Experiments in Coherent Decays of Neutral Mesons}
\author{Harry J. Lipkin}
\smallskip
\centerline{Department of Nuclear Physics}
\centerline{\it Weizmann Institute of Science}
\centerline{Rehovot 76100, Israel}

%\bk
\centerline{and}

%\bk
\centerline{School of Physics and Astronomy}
\centerline{Raymond and Beverly Sackler Faculty of Exact Sciences
}
\centerline{\it Tel Aviv University}
\centerline{Tel Aviv, Israel}
\centerline{January 11, 1995}
\abstract

Many theoretical calculations of subtle coherent effects in quantum mechanics
do not carefully consider the interface between their calculations and
experiment. Calculations for gedanken experiments using initial states
not satisfied in realistic experiments give results requiring
interpretation. Confusion and ambiguities frequently arise.
Calculations for time-dependent mixing
oscillations describe non-experiments. Physical experiments describe
oscillations in space in the laboratory system resulting from
interference between waves having
the same energy and time dependence; $not$ different momenta and space
dependence. Time-dependent oscillations are not observed.
\endpage

In recent kaon workshops, interesting theoretical questions arose
which are however not directly related to practical experimental
observables
\REF{\DafneHJL}{Harry J. Lipkin, Notes of a talk at a DA$\Phi$NE
Workshop, April 1993 (unpublished)}
$[{\DafneHJL}]$. Good theoretical calculations with very interesting
results are obtained from initial conditions which are not satisfied
in any experiment. The results are correct. But some kind of cook-book
recipe is needed to apply these results to the different conditions of a
real experiment.

Consider space and time dependence of neutral meson mixing. When a
neutral kaon is produced in an experiment as a $K^o$ or $\bar K^o$,
interference effects observed between the mass eigenstates $K_L$ and
$K_S$ can be interpreted to give a an experimental value for the mass
difference between these two states. However, confusion can arise if the
the initial state is not defined to correctly describe the experiment.

For example, in the reaction:
$$ K^- + p  \rightarrow \bar K^o + n                      \eqno (1) $$
a $K^-$ beam with a definite energy collides with a proton
target at rest, and the emitted neutron also has a definite energy.
Energy conservation requires the $\bar K^o$ to have a definite
energy. When it is split into the $K_L$ and $K_S$ components with
different masses, the two states have the same energy and different momenta.
Waves with different momenta propagate with different wave numbers;
their relative phase changes with distance. This gives an
oscillating interference pattern whose measurement gives the value of
the mass difference.

This experiment measures an interference between two kaon states
with the same $energy$ and different $momenta$, not interference between two
states with the same momenta and different energies. Similar effects have
been noted for neutrino oscillations
\REF{\Kayser}{Boris Kayser, Phys.Rev. D24 (1981) 110
}
\REF{\NeutHJL}{Harry J. Lipkin, Neutrino Oscillations and MSW
for Pedestrians, Lecture Notes (unpublished)}
$[{\Kayser},{\NeutHJL}]$.
However, most treatments of neutrino and neutral meson oscillations
describe oscillations in time resulting from waves having the same
momenta and different energies. These are ``non-experiments" which are
never performed in the laboratory and which require interpretation
for application to real experiments.

One can ask why one and not the other, since energy and momentum are
both conserved. We first note that if both the energies and momenta of
the initial state are known, then this is a ``missing mass experiment"
in which the mass of the outgoing kaon is determined uniquely by the
momentum of the outgoing neutron, and there will be no coherence nor
interference between the $K_L$ and $K_S$ waves. Coherence can occur only
when the uncertainty principle introduces a sufficient error in the
missing mass via the measurement of an observable which does not commute
with energy or momentum. In a real experiment, as opposed to a gedanken
non-experiment, position, not time is measured. In the reaction (1)
the position of the
proton in the initial state must be known with an error which is much
less than the wave length $\lambda$ of the interference pattern to be
measured. Thus there is an uncertainty in the momentum, which is
$$ \delta p >>  \hbar /\lambda                            \eqno(2a)  $$
This momentum uncertainty prevents a precise missing mass
measurement and allows coherence between the two outgoing kaon waves.
There is also an uncertainty in the kinetic energy
$$ \delta T  =  {{(\delta p)^2}/{2M_p}}                   \eqno(2b)  $$
Because the energy uncertainty is second order in the small quantity $\delta p$
it is negligible. Thus in a practical experiment, energy conservation can be
assumed, and the final state therefore has a definite energy. However, the
two kaon components must therefore have different momenta.

The difference in principle between energy and momentum coherence can be
stated as follows: If no explicit time measurement is made, the
components in the wave packet with different energies have different
frequencies and all relative phases average out in time. The explicit
measurement of a quantity which does not commute with momentum; namely
the positions of particles, measures a time-independent relative phase
between the components in the wave packet with different momenta.
The energy uncertainty (2b) can be taken into account
in a more precise calculation by expanding the initial wave function in
energy eigenstates, including the kinetic energy (2b) and
combining the results for different energy eigenstates incoherently.

The experiment is performed of course in the laboratory system.
A center-of-mass system cannot be defined for this
experiment. The center-of-mass system for the $n-K_L$ component of the final
state is not the same as the center-of-mass system for the $n-K_S$ component.
A theoretical treatment which begins with an initial state in the
center-of-mass system and has a final state in which the $K_L$ and $K_S$ have
the same momenta is describing a ``non-experiment". There is no simple way that
a coherent state of $K_L$ and $K_S$ can be produced with the same momenta
and different energies.

Some insight into the basic physics is obtained from a simple exercise
in quantum mechanics\REF{\LipQM}{Harry J. Lipkin, Quantum Mechanics,
North-Holland Publishing Co. Amsterdam (1973) pp.46-47 and 194-195}
$[{\LipQM}]$:
an experiment where a $K_L$ beam passes
through two regenerators separated by a distance $d$ and enters a $K_S$
detector. The question arises whether the two $K_S$ amplitudes arising
from the two regenerators are coherent.
Since the masses of the $K_L$ and $K_S$ are different, there must be an
energy or momentum transfer to the regenerator, and this can destroy the
coherence.

The answer is that the two amplitudes are coherent because of
a ``generalized M\"ossbauer effect". The two regenerators are not free;
they are bound to a table. The whole table takes up the recoil, providing
momentum transfer with negligible energy transfer. The $K_L$ and $K_S$
waves thus have the same energy and different momenta. The binding of the
regenerators to the table is crucial since the positions of the two
regenerators must known with an error much smaller than the wave
length to be measured. The regenerators are then bound in quantum states
whose uncertainty in momentum is very much larger than the momentum
transferred to the kaon beam by the regeneration. The kaon is thus
``elastically scattered" from the apparatus with finite momentum
transfer and negligible energy transfer.

The probability of this elastic scattering is given by the well-known
Debye-Waller factor $exp(-k^2\langle x^2\rangle)$ where k is the momentum
transfer and $\langle x^2\rangle$ is the mean square deviation of the
scatterer from its equilibrium position${\LipQM}]$. This probability is
very close
to unity for any realistic experiment where position fluctuations are
much smaller than the wave length of the oscillation to be measured. In
the general case this factor gives the probability of finding two
components in the momentum-space wave function which differs by $k$
in terms of the extension in configuration space of this wave function.
Thus in all experiments where spatial oscillations between two
neutral meson mass eigenstates are measured, the conditions required for
a feasible experiment insure that pairs of states with $different$
momenta and the $same$ energy must be present in the wave function and
these will give rise to a time-independent interference pattern in space.

We now show in a simple example how the description of a
time-dependent non-experiment can lead to ambiguities and confusion.
Consider $B - \bar B$ oscillations in one dimension where CP violation
and lifetime differences are neglected. The states
$\ket{B^o}$ and $\ket{\bar B^o}$ are equal mixtures with opposite
relative phase of the mass eigenstates denoted by
$\ket {B_L}$ and
$\ket{B_H}$ with masses denoted respectively by $M_L$ and $M_H$.
$$\ket{B^o} = (1/\sqrt 2)(\ket {B_L}+ \ket{B_H}); ~ ~ ~ ~
\ket{\bar B^o} = (1/\sqrt 2)(\ket {B_L} - \ket{B_H})
\eqno(3) $$

In an experiment where a $B^o$ is produced at x=0 in a state of
definite energy $E$, the momenta of the $B_L$ and $B_H$ components
denoted by $p_L$ and $p_H$ will be different and given by
$$p_L^2 = E^2 - M_L^2; ~ ~ ~ ~ ~ ~ p_H^2 = E^2 - M_H^2  \eqno(4) $$
Let $\ket{B^o(x)}$ denote this linear combination of $\ket {B_L}$ and
$\ket{B_H}$ with momenta $p_L$ and $p_H$ which is a pure $\ket{B^o}$
at $x=0$. The $\ket{B^o}$ and $\ket{\bar B^o}$ components of this wave
function will oscillate as a function of $x$ in a manner described by
the expression
$$
\left |
{{ \langle \bar B^o\ket{B^o(x)}}\over { \langle B^o\ket{B^o(x)}}}
\right |^2
=
\left |
{{e^{ip_L x} - e^{ip_H x}}\over {e^{ip_L x} + e^{ip_H x}}}
\right |^2
=
\tan^2 \left({{(p_L - p_H)x}\over{2}}\right) =
\tan^2 \left({{(M_L^2 - M_H^2)x}\over{2(p_L + p_H)}}\right)
\eqno(5) $$
These are just the normal $B - \bar B$ oscillations.

Now consider the ``non-experiment" often described in which a
a $B^o$ is produced at time t=0 in a state of
definite momentum $p$. The energies of the
$B_L$ and $B_H$ components
denoted by $E_L$ and $E_H$ will be different and given by
$$E_L^2 = p^2 + M_L^2; ~ ~ ~ ~ ~ ~ E_H^2 = p^2 + M_H^2  \eqno(6) $$
Let $\ket{B^o(t)}$ denote this linear combination of $\ket {B_L}$ and
$\ket{B_H}$ with energies $E_L$ and $E_H$ which is a pure $\ket{B^o}$
at $t=0$. The $\ket{B^o}$ and $\ket{\bar B^o}$ components of this wave
function will oscillate as a function of $t$ in a manner described by
the expression
$$ \left |{{ \langle \bar B^o\ket{B^o(t)}}\over
{ \langle B^o\ket{B^o(t)}}}\right |^2=
\left |{{e^{iE_L t} - e^{iE_H t}}\over {e^{iE_L t} + e^{iE_H t}}}\right |^2=
\tan^2 \left({{(E_L - E_H)t}\over{2}}\right) =
\tan^2 \left({{(M_L^2 - M_H^2)t}\over{2(E_L + E_H)}}\right)
\eqno(7) $$

In order to compare this result with a real experiment in which the
$B$ mesons are detected by a detector at a point $x$ it is necessary
to convert the gedanken time dependence into a real space dependence.
Here the troubles and ambiguities arise. One can simply convert time
into distance by using the relation
$$ x = vt = {{p}\over{E}} \cdot t                      \eqno(8a)  $$
where $v$ denotes the velocity of the $B$ meson. This immediately leads
to a result equivalent to the real experimental result (5), where the
small differences between $p_L$ and $p_H$ and between $E_L$ and $E_H$
are neglected.
$$ \left |{{ \langle \bar B^o\ket{B^o(t)}}\over
{ \langle B^o\ket{B^o(t)}}}\right |^2=
\tan^2 \left({{(M_L^2 - M_H^2)t}\over{2(E_L + E_H)}}\right) \approx
\tan^2 \left({{(M_L^2 - M_H^2)x}\over{4p}}\right)
\eqno(8b) $$

However, one can also argue that the $B_L$ and $B_H$ states with the
same momentum and different energies also have different velocities,
denoted by $v_L$ and $v_H$ and that they therefore arrive at the point
x at different times $t_L$ and $t_H$,
$$ x = v_Lt_L = {{p}\over{E_L}}\cdot t_L
= v_Ht_H = {{p}\over{E_H}}\cdot t_H                    \eqno(9a)  $$
One can then argue that the correct interpretation of the
time-dependent relation for measurements as a function of $x$ is
$$ \left |{{ \langle \bar B^o\ket{B^o(x)}}\over
{ \langle B^o\ket{B^o(x)}}}\right |^2=
\left |{{e^{iE_L t_L} - e^{iE_H t_H}}\over {e^{iE_L t_L} +
e^{iE_H t_L}}}\right |^2=
\tan^2 \left({{(E_Lt_L - E_Ht_H)}\over{2}}\right) =
\tan^2 \left({{(M_L^2 - M_H^2)x}\over{2p}}\right)
\eqno(9b) $$
This differs from the relations (5) and (8b) by a factor of 2 in the
oscillation wave length. If one does not consider the result of the
real experiment but only the two different interpretations of the
non-experiment, it is not obvious which of the two is correct.
There are also questions regarding whether phase velocity or group
velocity have been used in eqs.(8) and (9). All this confusion is
avoided by the direct use of the result (5) of the real experiment.

One can attempt to avoid the ambiguity and give a unique recipe for the
time-distance conversion by noting that the time constant of the
exponential decay of these unstable neutral mesons provides unambigous
time and length scales.
Let $t_e$ and $x_e$ denote
the points in time and space respectively where the initial state has
decayed by a factor of $e$ from its initial value,
$$ t_e = \tau \cdot {{E}\over {M}}; ~ ~ ~ ~ x_e = v t_e =
{{p}\over {E}} t_e = \tau \cdot {{p}\over {M}}     \eqno(10)     $$
where $\tau$ denotes the natural liftime of the relevant decay in the
rest system of the decaying state. This would be the $K_S$ lifetime for
kaon experiments and some mean lifetime for heavy quark states where the
lifetime differences between mass eigenstates are very small.
We neglect here small differences which are neglible in comparison with
the factor of 2 between eqs. (8b) and (9b) which remains to be
resolved. Substituting eqs. (10) into eqs. (5) and (7) gives
$$ \left |{{ \langle \bar B^o\ket{B^o(x_e)}}\over
{ \langle B^o\ket{B^o(x_e)}}}\right |^2=
\tan^2 \left({{(M_L^2 - M_H^2)x_e}\over{2(p_L + p_H)}}\right)
= \tan^2 \left({{(M_L - M_H) \tau}\over{2}} \cdot
{{(M_L + M_H)} \over{(p_L + p_H)}} \cdot{{p}\over{M}}\right) \approx
$$ $$ \approx
\tan^2 \left({{(M_L - M_H) \tau}\over{2}}\right)
\eqno(11a) $$
$$ \left |{{ \langle \bar B^o\ket{B^o(t_e)}}\over
{ \langle B^o\ket{B^o(t_e)}}}\right |^2=
\tan^2 \left({{(M_L^2 - M_H^2)t_e}\over{2(E_L + E_H)}}\right)
= \tan^2 \left({{(M_L - M_H) \tau}\over{2}} \cdot
{{(M_L + M_H)} \over{(E_L + E_H)}} \cdot{{E}\over{M}}\right) \approx
$$ $$ \approx
\tan^2 \left({{(M_L - M_H) \tau}\over{2}}\right)
\eqno(11b) $$
where we have set $p$, $M$, and $E$ to the mean of the values for the two
eigenstates. The results (11) agree with eq.(8b) and disagree with
eq.(9b). They are also confirmed by the simple case of a decay at
rest, where the phase is clearly $(M_L - M_H) \tau/2 $ for the case where
the amplitude has decayed exponentially by a factor $e$. The results
(11) have a simple and clear physical interpretation. They give the
ratio between the imaginary and real parts of the eigenvalues of the mass
matrix. These are the same for both the gedanken time-dependent
experiment and the real space-dependent experiment. The argument leading
to eq. (9b) does not have a well-defined meaning in terms of a definite
experiment, either real or gedanken. It attempts to use the results
obtained for the initial state of the gedanken experiment in the geometry
of the real experiment. This evidently involves some double counting to
account for the factor of two.

We immediately note the analogous implications for all
experiments measuring $B - \bar B$ oscillations.
Calculations for $B - \bar B$ oscillations
in time describe non-experiments. Times are never measured in the
laboratory; distances are measured. When correlated decays of two mesons
will be measured in an asymmetric B factory, the points in space where
the two decays will be measured in the laboratory, not the time
difference which appears in many calculations.

If the points in space where two correlated B decays occur are measured
with sufficient precision to describe meaningful oscillations, the above
discussion shows that this precision in position introduces a crucial
momentum uncertainty. Calculations describing the real experiment
directly are most simply performed in the laboratory system where
interference occurs between waves having the same energy and different
momenta. Waves with different energy are not coherent and cannot
interfere if there is no explicit time measurement.

Lorentz boosts are essentially useless; they mix uncertainties in
momentum and energy and oscillations in space and time. Just as there is
no unique definition of a center-of-mass system for the kaon experiment,
there are no unique definitions of the center-of-mass system for the two
$B$ mesons, of the rest system of a given $B$ meson or of a proper time
for the observed decay. However, everything can be described simply in
the laboratory, where the energy eigenstates can be well defined,
different waves with the same energy are coherent and waves with
different energies are incoherent.

This confusion between distance and time measurements
does not arise in measurements of single waves rather than interference
between two nearly degenerate waves. There is no such ambiguity in
experiments using a distance measurement to measure
the lifetime of a mass eigenstate with a well defined velocity. The
exponential decay in time in the rest frame of the mass eigenstate is
easily transformed to an exponential in space in the laboratory. Here
the uncertainties in momentum introduced by the uncertainty principle are
negligible. This clearly applies to the kaon states where the mass
eigenstates are easily separated. It also applies to $D$ and $B$ mesons
in the approximation where the lifetime difference between the two mass
eigenstates is neglected and a single exponential is measured.

An oscillation frequency for neutral mixing is measured without a time
dependence in experiments where the time integral of the oscillations
is measured; e.g. when the decays of a state tagged initially as a $B^o$
oscillates between $B^o$ and $\bar B^o$ and the relative numbers of
$B^o$ and $\bar B^o$ measured are effectively summed over space and time.
This gives the ratio of the oscillation frequency to the exponential
decay rate; i.e. the ratio of the imaginary and real parts of the mass
matrix eigenvalues and is always the same in both space and time.

This research is supported in part by grant no. I-0304-120.07/93
from the German-Israeli Foundation for Scientific Research and
Development.

\refout
\end